# Microscopic Origins of Electron and Hole Stability in ZnO

C. Richard A. Catlow, Alexey A. Sokol, and Aron Walsh

University College London, Kathleen Lonsdale Materials Chemistry, Department of Chemistry, 20 Gordon Street, London WC1H 0AJ, UK

http://www.ucl.ac.uk/klmc

**We present a fundamental method to assess the doping limits of hetero-polar materials; applied to the case of ZnO, we show clearly that electrons are stable and holes are unstable under the limits of thermodynamic control.**

Metal oxides of intermediate band gap are extensively studied owing to their wide-ranging actual and potential applications in electronic and opto-electronic devices[1]. A crucially important factor in such applications is the extent to which the electronic structure may be controlled by doping. A widely discussed example concerns the possibility of acceptor, or "*p*-doping" of ZnO, which has largely proved to be elusive[2]. Indeed, it is recognised that fundamental factors relating to the electronic structure of the materials limit the extent to which the electronic structure may be modified by doping[3-8]. Here, we examine this problem using a microscopic approach in which we calculate the energies associated with the basic defect reactions controlling hole, electron and defect concentrations. Our approach allows us to assess straightforwardly whether doping will lead to the introduction of electronic or defect states. We apply our method to the case of ZnO. Theoretical and experimental studies of holes localised on the oxygen sublattice in oxides have been recently reviewed by Schirmer[9]; there is ample evidence for the presence of bound hole states in ZnO[10]. Initial experimental reports of *p*-type conduction in ZnO[11-13] and a theoretical proposal of the co-doping approach to the material fabrication[14] have been followed by an exponentially growing number of studies[15-17a]. However, our results suggest that *p*-doping, with free carrier formation, may not be feasible under conditions of thermodynamic control in the bulk material.

---

[a] *A current search on Web of Knowledge reveals in excess of 1,700 publications on this topic in the last decade.*

Modification of the concentrations of electronic species (holes or electrons) in oxides may be achieved by changes in stoichiometry or by the introduction of aliovalent dopants, for which hole or electron species are introduced as charge compensators. For example, monovalent cation dopants (*e.g.* Li$^+$) substituting at a cation site in a divalent oxide, such as ZnO, have an effective negative charge and may be charge compensated by a hole; while doping with a monovalent anion dopant at the anion site gives a species with an effective positive charge leading to electron compensation. However, the alternative compensation by point defects is always in principle possible. Thus in the case of ZnO, dopants with an effective negative charge may be compensated by oxygen vacancies or zinc interstitials and those with an effective positive charge by zinc vacancies or oxygen interstitials. Which of these modes dominates depends on basic defect thermodynamics; and in every case we may formulate a redox reaction whereby electronic species may be exchanged with point defect compensation. Let us take, for example, acceptor or "*p*-doping" of ZnO: an electron hole is converted to oxygen vacancy compensation by the redox reaction[b]:

(1) $$h^\bullet + \frac{1}{2}O_O^x \rightleftharpoons \frac{1}{2}V_O^{\bullet\bullet} + \frac{1}{4}O_2(g)$$

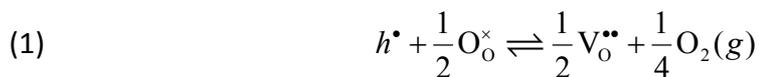

If the energy of such a reaction is appreciably negative, then holes will be thermodynamically unstable with respect to vacancies under normal oxygen partial pressures. Corresponding reactions may be written for the interchange of hole with zinc interstitial compensation. In addition, redox processes may be formulated involving metal rather than oxygen exchange; and although the latter is more likely under normal operating conditions, we have formulated all possible defect redox reactions for the redox controlled interconversion of electronic with point defect compensation in binary divalent oxides. These reactions are summarised and collected in Table (1a) for the case of ZnO.

Calculations of the energies of these reactions will allow us to predict whether hole or electron doping is possible. Fortunately, the techniques for calculating the formation energies of electronic and defect states in oxides are well developed. We are able, therefore, to illustrate our approach for the case of ZnO. Earlier work[18,19] has calculated the necessary energies using the hybrid quantum-mechanical/molecular-mechanics (QM/MM) approach, where an embedded quantum mechanical cluster, containing the defect or electronic state, is described using Density Functional Theory (the hybrid non-local B97-1 functional, which improves upon commonly used local or semi-local exchange and correlation functionals) and the embedding molecular mechanics region with a carefully parameterised Born Shell Model potential. Importantly, this approach avoids artificial interactions between periodic images of point defects, while providing a proper dielectric and elastic response to point defect formation, and an unambiguous energy reference (vacuum level) from which ionisation energies can be calculated; these issues have been tackled, but are not yet fully resolved for periodic supercell calculations. Such methods have achieved good agreement with experiment and are, we consider, sufficiently accurate for our present purposes, although we return to this question below.

---

[b] The defect reactions are formulated using the standard Kröger-Vink notation.

Table (2) collects the component terms needed to calculate the energies of the redox reactions presented in Table (1a) for the case of ZnO. The resulting energies, given in the same table, show that for acceptor doped materials, defect compensation is in all cases energetically more favourable than compensation by holes, *suggesting that except under conditions of exceptionally high oxygen chemical potential, p-doped material will not be achievable under conditions of thermodynamic equilibrium* – a result that is consistent with the known difficulty of effective acceptor doping of ZnO. Of course holes may be stabilised by trapping by their charge compensating dopant; but only free, untrapped holes, to which our calculations refer, can act as charge carriers.

We have also applied the same analysis to the case of donor or electron doped material (*i.e.* those in which the dopant has a positive effective charge). The relevant redox reactions are given in Table (1b) together with the resulting calculated energies for the case of ZnO. In this case we find that electron is strongly favoured over defect compensation, which clearly accords with the known efficacy of donor doping in ZnO.

How reliable are our calculated energies? As is invariably the case when estimating energies of thermodynamic cycles, the resulting energy is relatively small in magnitude compared with the component terms. The energies of the holes/electrons and defects are clearly crucial. Regarding the former, we note the close agreement of our calculated ionisation potential of 7.71 eV with the experimental value of *ca.* 7.82 eV[20]. (In this context we note that our model for the hole state assumes a delocalised, non-polaronic state; calculations of the relaxation energy associated with hole localisation estimate an energy of ~0.6 eV – far less than half the valence band width (~5.3 eV)[2] of ZnO, suggesting that the delocalised model is valid[21,22].) Regarding the point defect energies, it is not possible to make direct comparison with experiment; but we note that reference 19 also calculated these energies using the classical Mott-Littleton method[23] and that there was close quantitative agreement between the energies calculated by the two different methods, which gives us confidence in the reliability of both. To provide further guidance and insight, we have calculated the energy of reaction (1) using a semi-classical approach based on a Born-Haber cycle employing Mott-Littleton energies, details of which will be given elsewhere. Such cycles require estimates of the second electron affinity of oxygen concerning which there is some uncertainty[24,25]. The resulting calculated energy of -1.4 eV for the reaction is larger than that calculated using the DFT based methods, but again indicates thermodynamic instability of holes with respect to oxide vacancies. Definitive values for these energies would be of great value as they would indicate the extent to which defect levels must be shifted, by *e.g.* lattice strain, if electronic rather than point defect doping is to be achieved.

The results summarised in this letter suggest that the known difficulty in *p*-doping of ZnO has a fundamental thermodynamic basis rooted in the defect and electronic energetic in the material. Of equal significance is the approach we have formulated, which enables the calculation, by widely available techniques, of the energies of the defect reactions that control the extent of electronic versus defect concentrations. Analyses of other oxide materials including $SnO_2$ and $In_2O_3$ will be presented shortly.

**Acknowledgements**

The work has been supported by an EPSRC Portfolio Partnership (Grant No. ED/D504872) and membership of the UK's HPC Materials Chemistry Consortium, which is funded by EPSRC (Grant No. EP/F067496). A.W. would like to acknowledge funding from a Marie-Curie Intra-European Fellowship from the European Union under the Seventh Framework Programme.

**Table 1**. Reaction energies (ΔE$_f$), as calculated using an embedded cluster approach, for defect processes in ZnO, in which electron and hole carriers are charge compensated by ionic defects. Negative values refer to exothermic reactions (all hole compensation processes), while positive values refer to endothermic reactions (all electron compensation processes). The defect reactions are formulated using the standard Kröger-Vink notation, where a superscript prime refers to an extra negative charge, while a dot refers to an extra positive charge.

|  | (a) Hole Carriers | | (b) Electron Carriers | |
| --- | --- | --- | --- | --- |
|  | Defect Reaction | ΔE$_f$ (eV) | Defect Reaction | ΔE$_f$ (eV) |
| **Oxygen rich** | $h^\bullet + \frac{1}{2}ZnO(s) \rightleftharpoons \frac{1}{2}Zn_i^{\bullet\bullet} + \frac{1}{4}O_2(g)$ | -0.50 | $e' + \frac{1}{4}O_2(g) \rightleftharpoons \frac{1}{2}O_i''$ | 1.73 |
| | $h^\bullet + \frac{1}{2}O_O^\times \rightleftharpoons \frac{1}{2}V_O^{\bullet\bullet} + \frac{1}{4}O_2(g)$ | -0.74 | $e' + \frac{1}{2}Zn_{Zn}^\times + \frac{1}{4}O_2(g) \rightleftharpoons \frac{1}{2}V_{Zn}'' + \frac{1}{2}ZnO(s)$ | 0.75 |
| **Zinc rich** | $h^\bullet + \frac{1}{2}O_O^\times + \frac{1}{2}Zn(s) \rightleftharpoons \frac{1}{2}V_O^{\bullet\bullet} + \frac{1}{2}ZnO(s)$ | -2.59 | $e' + \frac{1}{2}Zn_{Zn}^\times \rightleftharpoons \frac{1}{2}V_{Zn}'' + \frac{1}{2}Zn(s)$ | 2.60 |
| | $h^\bullet + \frac{1}{2}Zn(s) \rightleftharpoons \frac{1}{2}Zn_i^{\bullet\bullet}$ | -2.35 | $e' + \frac{1}{2}ZnO(s) \rightleftharpoons \frac{1}{2}O_i'' + \frac{1}{2}Zn(s)$ | 3.58 |

**Table 2**. Energies of component electronic and defect species in ZnO[19]. The defect reactions are formulated using the standard Kröger-Vink notation.

| Species | Fundamental Defect Reactions | Energy (eV) |
|---|---|---|
| $V_{Zn}^{//}$ | $Zn_{Zn}^{x} \rightleftharpoons V_{Zn}^{//} + 2h^{\bullet} + Zn(s)$ | 12.077 |
| $V_{O}^{\bullet\bullet}$ | $O_{O}^{x} \rightleftharpoons V_{O}^{\bullet\bullet} + 2e^{/} + \frac{1}{2}O_{2}(g)$ | 5.397 |
| $O_{i}^{//}$ | $\frac{1}{2}O_{2}(g) \rightleftharpoons O_{i}^{//} + 2h^{\bullet}$ | 10.345 |
| $Zn_{i}^{\bullet\bullet}$ | $Zn(s) \rightleftharpoons Zn_{i}^{\bullet\bullet} + 2e^{/}$ | 2.177 |
| $h^{\bullet}$ | $0 \rightleftharpoons h^{\bullet} + e_{vacuum}^{/}$ | 7.714 |
| $e^{/}$ | $e_{vacuum}^{/} \rightleftharpoons e^{/}$ | -4.277[a] |
| ZnO | $Zn(s) + \frac{1}{2}O_{2}(g) \rightleftharpoons ZnO(s)$ | -3.704[b] |

[a]The electron energy is taken as the sum of the calculated ionisation potential (hole energy) and the electronic band gap of 3.437 eV[2].

[b]For the bulk material, we use the enthalpy of formation at 0 K with respect to the elemental standard states[26].